\documentclass[12pt,twoside]{article}
\usepackage{fleqn,espcrc1,epsfig}

\def\rsim{\mathrel{\raise2pt\hbox to 8pt{\raise -5pt\hbox{$\sim$}\hss{$>$}}}}
\def\lsim{\mathrel{\raise2pt\hbox to 8pt{\raise -5pt\hbox{$\sim$}\hss{$<$}}}}

\begin {document}

\title{Electromagnetic Structure of Few-Nucleon Systems: a Critical Review}
\author{R.\ Schiavilla
\address{Jefferson Lab, Newport News, Virginia 23606 \\
         $\>\>$and \\
         $\>\>$Department of Physics, Old Dominion University,
         Norfolk, Virginia 23529}}
\maketitle

\begin{abstract}
Our current understanding of the structure of
nuclei with $A \leq 8$, including energy spectra, electromagnetic
form factors, and capture reactions, is critically reviewed
within the context of a realistic approach to nuclear dynamics
based on two- and three-nucleon interactions and associated
electromagnetic currents.
\end{abstract}

\section{Introduction}

In the present talk I will review the simple, traditional
picture of the nucleus as a system of point-like nucleons
interacting among themselves via effective many-body potentials,
and with external electro-weak probes via effective many-body
currents.  I will also discuss the extent to which this
picture, the so called \lq\lq nuclear standard model\rq\rq,
is successful in predicting a number of nuclear properties
for systems with mass number $A\leq 8$, including energy spectra
of low-lying states, electromagnetic form factors, and low-energy
capture reactions.
\section{Interactions and Energy Spectra}

The Hamiltonian in the nuclear standard model is taken to
consist of a non-relativistic kinetic energy operator, 
and two- and three-nucleon potentials.
The two-nucleon potential consists of a long range
part due to pion exchange, and a short-range part parameterized
either in terms of heavy meson exchanges as, for example, in
the Bonn potential~\cite{Bonn}, or via suitable operators
and strength functions, as in the Argonne $v_{18}$
(AV18) potential~\cite{WSS95}.  The short-range terms in
these potentials are then constrained to fit $pp$ and $np$
scattering data up to energies of $\simeq$ 350 MeV in the
laboratory, and the deuteron binding energy.  The modern
models mentioned above include isospin-symmetry-breaking
components, and provide fits to the Nijmegen
data-base~\cite{Nijm} characterized by $\chi^2$ per datum
close to one.  They should therefore be viewed as phase-equivalent.

\begin{figure}[bth]
\centerline{
\epsfig{file=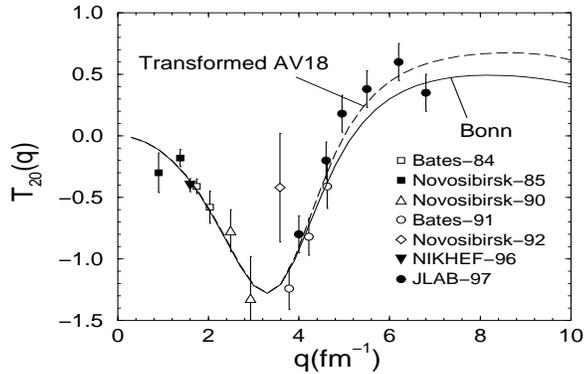,height=2in,width=3in}}
\caption{The deuteron tensor polarizations obtained with
Bonn potential and \lq\lq unitarily transformed\rq\rq
AV18 deuteron wave function and charge operators.} 
\label{fig:t20}
\end{figure}

A major difference among these modern potential models, however,
is in the treatment of non-localities, particularly those
associated with off-the-energy-shell prescriptions of the
one-pion-exchange (OPE) term.  The AV18 as well as the Nijmegen
models incorporate the on-shell form of the OPE term,
which leads to a local tensor component.  The Bonn potential, on
the other hand, includes the off-shell extension predicted
by pseudo-scalar coupling of pions to nucleons, and hence has
a strongly non-local tensor component.  More than two decades
ago, Friar and collaborators~\cite{Fri77} showed that different OPE off-shell
extensions can be related to each other via a unitary
transformation, and that differences in predictions
for observables sensitive to OPE, such as the triton binding
energy or the deuteron tensor polarization at moderate momentum
transfers ($\leq$ 5 fm$^{-1}$), can be, to a large
extent, removed by performing {\it consistent} calculations,
namely calculations using three-nucleon interactions and charge
operators that obey the unitary equivalence of the OPE interaction.
An example of this type of calculations is illustrated in
Fig.~\ref{fig:t20}, where the deuteron tensor polarizations
corresponding to the Bonn potential and to the AV18 deuteron wave
function and charge operator, unitarily transformed to match the
off-shell extension of the Bonn OPE, are compared.  The remaining
differences at the larger values of momentum transfer are presumably
originating from additional short-range non-localities present in the Bonn
model.  It should be stressed that {\it consistent} calculations
of the type alluded to above have been carried out up until now
only for the deuteron.  It would be interesting to verify these
expectations also for the case of other observables, such as the triton
binding energy, for example.

It is now well established that two-nucleon potentials alone underbind
nuclei~\cite{Wir00,NKG00}: for example, the AV18 and Bonn models give~\cite{NKG00}, in
numerically exact calculations, binding energies of 24.28 MeV and 26.26 MeV
respectively, which should be compared to the experimental value
of 28.3 MeV.  Moreover, it has been shown in Ref.~\cite{Wir00} that, for example, $^6$Li
and $^7$Li are unstable against breakup into $\alpha$$d$ and
$\alpha$$t$ clusters, respectively, and that energy differences are not, in
general, well predicted, when only two-nucleon potentials are retained
in the Hamiltonian.

Important components of the three-nucleon potential
are due to the internal structure of the nucleon.
Since all degrees of freedom other than the nucleon
have been integrated out, the presence of virtual $\Delta$
resonances, for example, induces three-nucleon potentials.
Current three-nucleon potential models are reviewed by
Carlson~\cite{Car00} in these proceedings.
These newly developed models include the \lq\lq long-range\rq\rq term, resulting from
the intermediate excitation of a $\Delta$ with pion exchanges involving the
other two nucleons, known as the Fujita-Miyazawa term~\cite{FM57},
as well as multipion exchange terms
involving excitation of one or two $\Delta$'s, so-called pion-ring
diagrams, and the terms arising from S-wave pion rescattering,
required by chiral symmetry.  There are four strength parameters
which are then determined by fitting the energies of $\simeq 20$  
low-lying states of systems with $A \leq 8$ nuclei in exact
Green's function Monte Carlo calculations.  The resulting
energy spectra for these systems are in good agreement with
the experimental ones~\cite{Car00}.  In particular, the underbinding
of neutron-rich nuclei, such as $^6$He and $^8$Li, which had
proven to be a problem with earlier models of three-nucleon
potentials, is resolved to a large extent.

\section{Electromagnetic Current and Form Factors}

The nuclear current operator consists of one- and many-body terms
that operate on the nucleon degrees of freedom.  The one-body
operator has the standard expression in terms of single-nucleon
convection and magnetization currents.  The two-body current
operator has \lq\lq model-independent\rq\rq and
\lq\lq model-dependent\rq\rq components (for a review, see Ref.~\cite{Car98}).
The model-independent terms are obtained from the two-nucleon 
potential, and by construction satisfy current conservation
with it.  The leading operator is the isovector
\lq\lq $\pi$-like\rq\rq current obtained
from the isospin-dependent spin-spin and tensor interactions.
The latter also generate an isovector \lq\lq $\rho$-like \rq\rq current, while
additional model-independent isoscalar and isovector currents arise from the
central and momentum-dependent interactions.  These currents are short-ranged
and numerically far less important than the $\pi$-like current.  Finally,
models for three-body currents have been derived in Ref.~\cite{Mar98}, by
gauging the two-pion exchange three-nucleon interaction associated with S-wave
pion-nucleon scattering.  The resulting contributions have been found to
be very small in studies of the magnetic structure of the trinucleons~\cite{Mar98}.

The model-dependent currents are purely transverse
and therefore cannot be directly linked to the underlying
two-nucleon interaction.  Among them, those associated with
the $\Delta$-isobar are the most important ones at moderate
values of momentum-transfer ($q \leq 5$ fm$^{-1}$).  These
currents are treated within the transition-correlation-operator
scheme~\cite{Mar98,Sch92}, a scaled-down
approach to a full $N$+$\Delta$ coupled-channel treatment.
In this scheme, the $\Delta$ degrees of freedom
are explicitly included in the nuclear wave functions by
means of transition correlation operators
that convert $NN$ pairs into $N\Delta$ and $\Delta\Delta$ pairs, acting
on a purely nucleonic wave function.  Both $\gamma N \Delta$ and $\gamma \Delta \Delta$
$M_1$ couplings are considered with their values obtained from data~\cite{Sch92}.

\begin{figure}[bth]
\centerline{
\epsfig{file=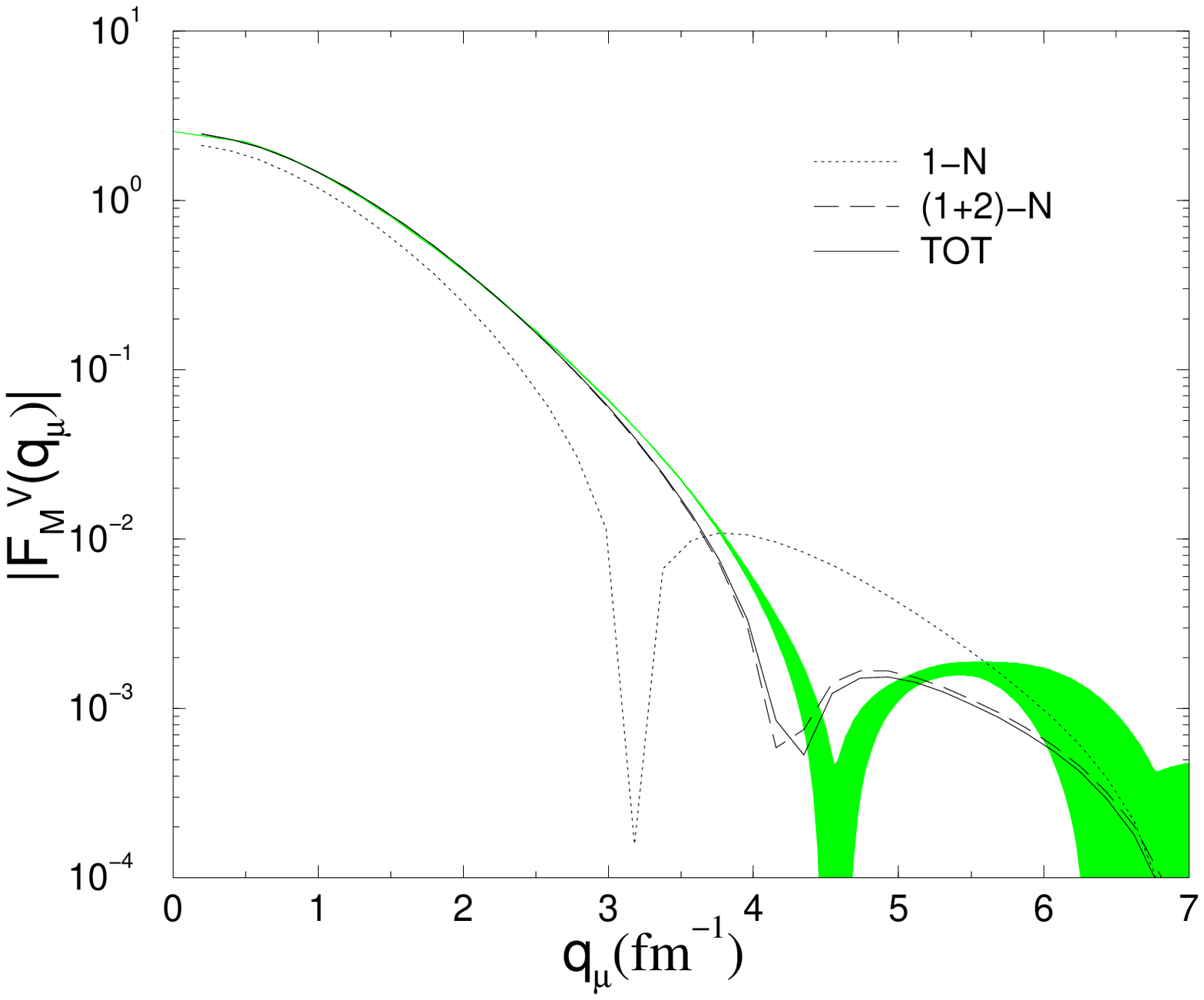,height=3in,width=3in}
\epsfig{file=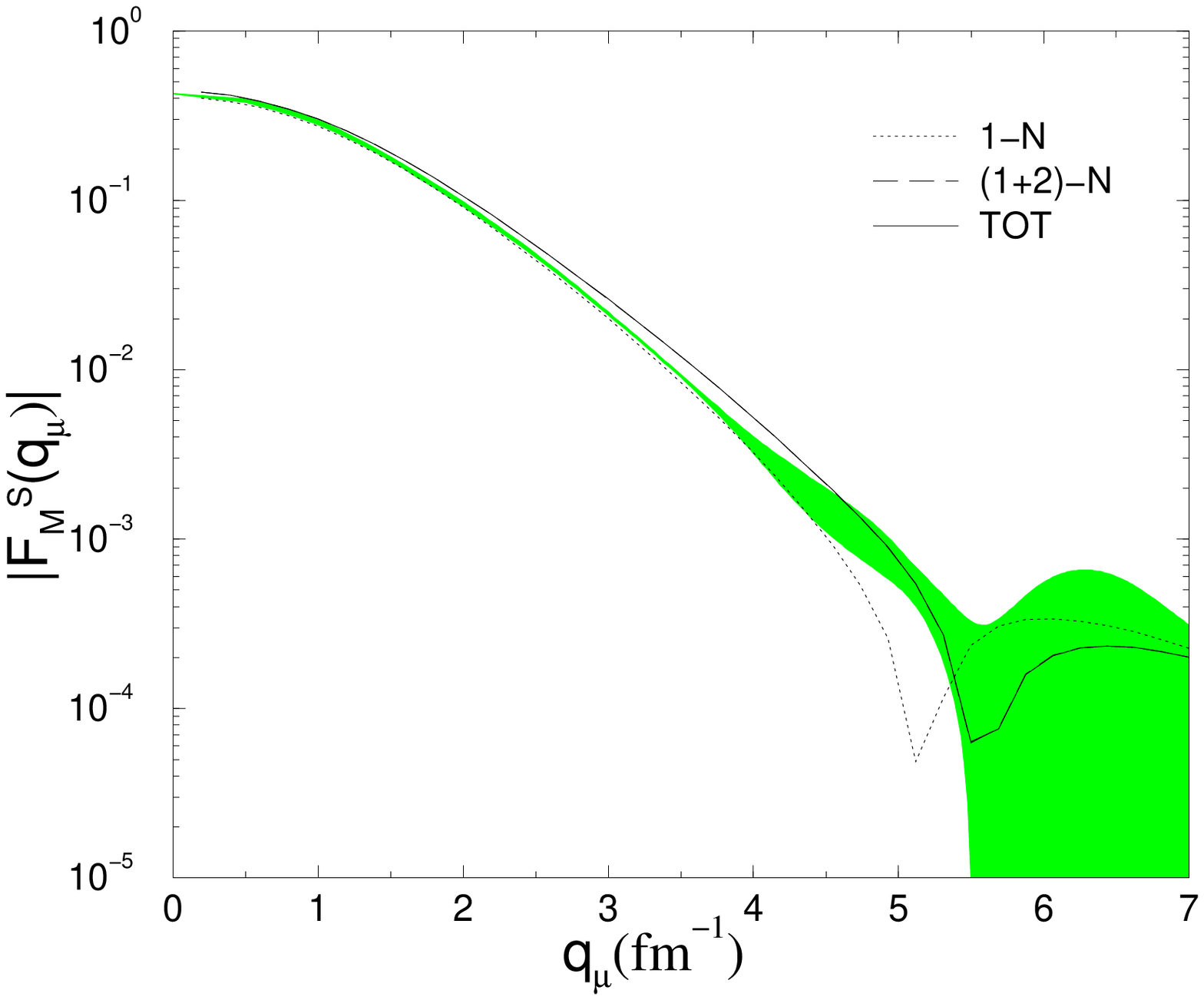,height=3in,width=3in}}
\caption{The isovector and isoscalar combinations
of the $^3$H and $^3$He magnetic form factors obtained
with the AV18/UIX Hamiltonian.  The contributions
associated with nucleonic one-body and (one+two)-body, and
$\Delta$ currents are displayed.}
\label{fig:ffm3}
\end{figure}

The calculated isoscalar and isovector magnetic form factors
of the trinucleons are shown in Fig.~\ref{fig:ffm3}.
The isovector form factor is undepredicted by theory in the
first diffraction region.  In this region, the $\pi$-like and $\rho$-like currents,
constructed from the spin-spin and tensor components of the $v_{18}$
interaction in the results shown in Fig.~\ref{fig:ffm3}, is the dominant
contribution, and therefore the underprediction mentioned above indicates
that these currents are too weak at moderate values of momentum transfers.
On the other hand, the isovector magnetic moment is in excellent
agreement with the experimental value.

The isoscalar form factor appears to be slightly overpredicted by
theory over the whole momentum transfer range.  In particular, the
calculated isoscalar magnetic moment is roughly 4 \% too large with
respect to the experimental value.  This again points to deficiencies
in the model for two-body currents.

While the model-independent two-body currents are linked to the form of
nucleon-nucleon interaction
via the continuity equation, the most important two-body charge operators
are model dependent and may be viewed as relativistic corrections.
They fall into two classes.  The first class includes those effective
operators that represent non-nucleonic degrees of freedom,
such as nucleon-antinucleon pairs or nucleon-resonances, and which arise when
these degrees of freedom are eliminated from the state vector.  To the
second class belong those dynamical exchange charge effects that would
appear even in a description explicitly including non-nucleonic excitations
in the state vector, such as the $\rho\pi\gamma$
transition coupling.  The proper forms of the former operators depend on
the method of eliminating the non-nucleonic degrees of
freedom~\cite{Fri77}.  There are nevertheless rather clear indications
for the relevance of two-body charge operators from the failure of calculations based on
the one-body operator in predicting the charge
form factors of the three- and four-nucleon systems,
and deuteron tensor polarization observable.

The largest two-body charge contribution, at moderate values
of the momentum transfer, is that due to the $\pi$-meson
exchange operator.  It is derived by considering the low-energy limit of the
relativistic Born diagrams associated with the virtual $\pi$-meson
photoproduction amplitude.  In this limit, such an amplitude leads
to two terms (at the lowest order).  The first term consists of
three factors: a single-nucleon charge operator,
a non-relativistic propagator, and a OPE interaction.
It is already included in the one-body (impulse approximation)
calculation of the form factors, since the wave functions used in
these calculations are obtained from solutions of a Schr\"odinger
equation including the OPE interaction.  The second term, however,
represents the truly two-body charge operator.  It is a local operator
with both isoscalar and isovector components.
There are additional contributions due to the energy
dependence of the pion propagator and direct coupling of the photon to
the exchanged pion~\cite{Fri77}.  These operators, however, give rise
to non-local isovector contributions which are expected to provide only
small corrections to the leading local terms.  For example these operators
would only contribute to the isovector combination of the $^{3}$He and
$^{3}$H charge form factors, which is anyway a factor of three smaller
than the isoscalar.  Thus they have been neglected in most of the studies I am
familiar with.

\begin{figure}[bth]
\centerline{
\epsfig{file=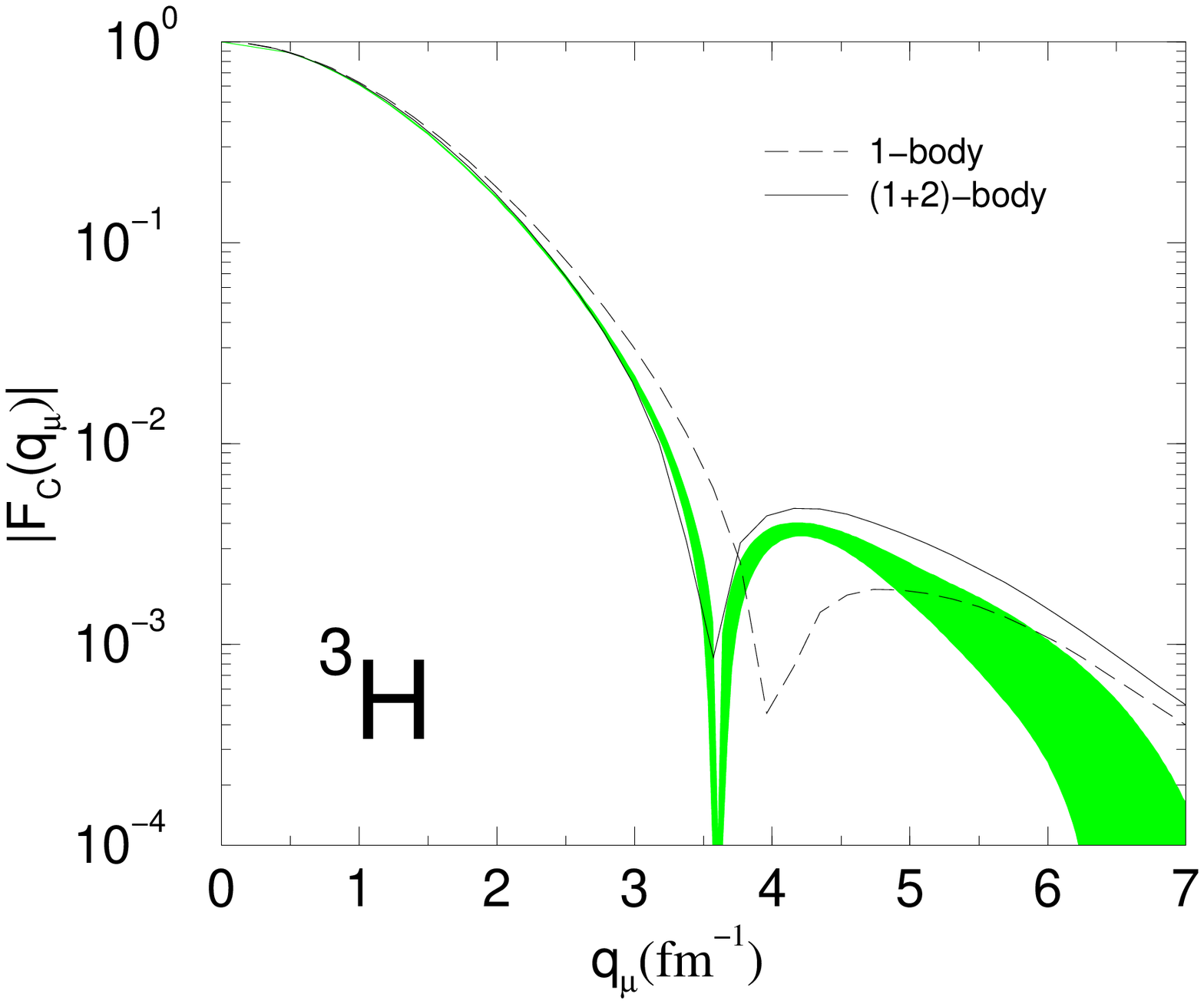,height=3in,width=3in}
\epsfig{file=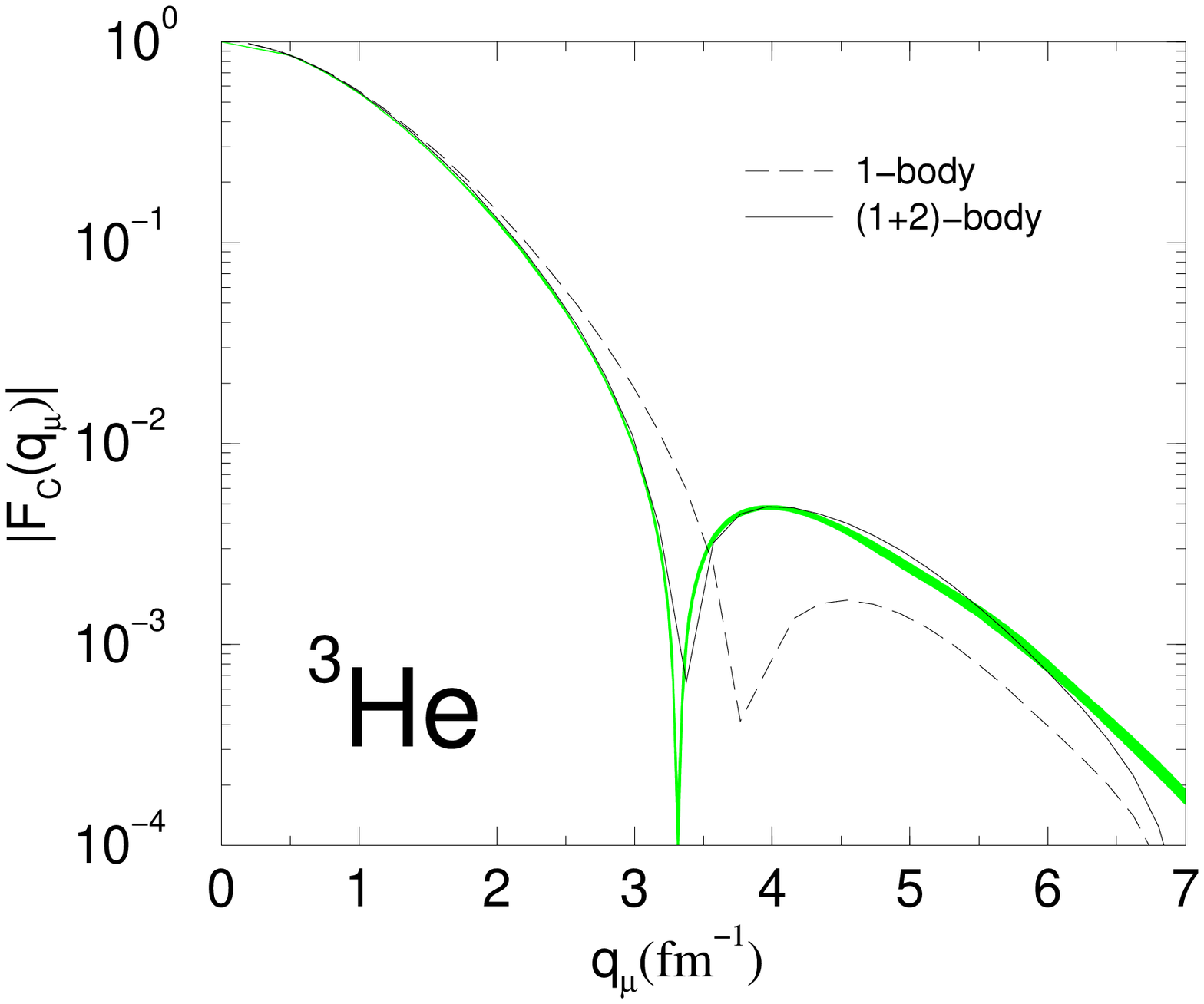,height=3in,width=3in}}
\caption{The $^3$H and $^3$He charge form factors obtained
with the AV18/UIX Hamiltonian.  The contributions
associated with one-body and (one+two)-body operators are displayed.}
\label{fig:ffc3}
\end{figure}

The calculated $^3$He and $^3$H charge form factors are compared
to data in Fig.~\ref{fig:ffc3}.  There is excellent agreement between theory and
experiment.  The important role of the two-body contributions
above 3 fm$^{-1}$ is also evident.  The remarkable success of
the present picture based on non-relativistic wave functions
and a charge operator including the leading relativistic
corrections should be stressed.  It suggests, in particular, that the
present model for the two-body charge operator is better than one
{\it a priori} should expect.  These operators
fall into the class of relativistic corrections.  Thus,
evaluating their matrix elements with non-relativistic wave functions
represents only the first approximation to a systematic reduction.  A
consistent treatment of these relativistic effects would require, for
example, inclusion of the boost corrections on the nuclear wave
functions.  Yet, the excellent agreement between the calculated
and measured charge form factors suggests that these corrections
may be neglegible in the $q$-range explored so far.

\section{Capture Reactions}

\subsection{The $pd$ and $nd$ Radiative Captures}

There are now available many high-quality data, including differential
cross sections, vector and tensor analyzing powers, and photon polarization
coefficients, on the $pd$ radiative capture at c.m.\ energies
ranging from 0 to 2 MeV~\cite{Sea96,Mea97,Wea99,SK99}.  These data indicate
that the reaction proceeds predominantly through S- and P-wave capture.
\begin{figure}[bth]
\centerline{
\epsfig{file=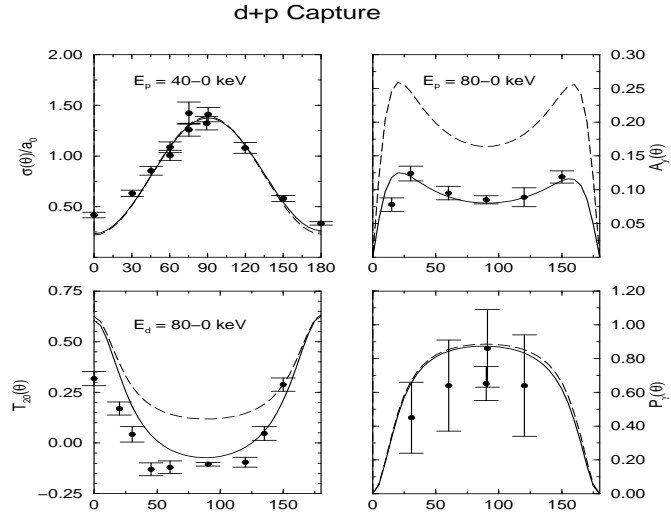,height=2.5in,width=3.5in}}
\caption{The energy integrated cross section $\sigma(\theta)/a_0$
($4\pi a_0$ is the total cross section), vector analyzing power
$A_y(\theta)$, tensor analyzing power $T_{20}(\theta)$ and photon
linear polarization coefficient $P_\gamma(\theta)$ obtained with the
AV18/UIX Hamiltonian model and one-body only (dashed line)
or both one- and many-body (solid line) currents
are compared with the experimental results of Ref.~\protect\cite{Sea96}.}
\label{fig:cpt080keV}
\end{figure}

The predicted angular distributions~\cite{Viv00} of the differential
cross section $\sigma(\theta)$, vector and tensor analyzing powers
$A_y(\theta)$ and $T_{20}(\theta)$, and photon
linear polarization coefficient $P_\gamma(\theta)$ are compared with
the TUNL data below 50 keV from
Refs.~\cite{Sea96,Wea99} in Fig~\ref{fig:cpt080keV}.  Note that
the AV18/UIX Hamiltonian model is used in the calculations reported here.
The agreement between the full theory, including many-body
current contributions, and experiment is generally good.
However, a closer inspection of the figure reveals the
presence of significative discrepancies between theory and experiment
in the small angle behavior of $\sigma(\theta)$ and $T_{20}(\theta)$, as well
as in the $S$-factor below 40 keV~\cite{Viv00}.
The S-wave capture proceeds mostly through the $M_1$ transitions connecting
the doublet and quartet $pd$ states to $^3$He--the associated reduced
matrix elements (RMEs) are denoted by $m_2$ and $m_4$, respectively.
The situation for P-wave capture is more complex, although at energies
below 50 keV it is dominated by the $E_1$ transitions
from the doublet and quartet $pd$ states having channel spin
$S$=1/2, whose RMEs I denote as $p_2$ and $p_4$.  The $E_1$
transitions involving the channel spin $S=3/2$ states, while
smaller, do play an important role in $T_{20}(\theta)$.

The TUNL~\cite{Wea99} and Wisconsin~\cite{SK99} groups have determined
the leading $M_1$ and $E_1$ RMEs via fits to the measured observables.
The results of this fitting procedure are compared with the
calculated RMEs in Table~\ref{tab:rmew}.
The phase of each RME is simply related to
the elastic $pd$ phase shift~\cite{SK99}, which at these low energies
is essentially the Coulomb phase shift.
As can be seen from Table~\ref{tab:rmew},
the most significant differences between theoretical and experimental RMEs
are found for $|p_4|$.  The theoretical overprediction of $p_4$
is the cause of the discrepancies mentioned above in the
low-energy ($\le 50$ keV) $S$-factor and small
angle $\sigma(\theta)$.

It is interesting to analyze the ratio
$r_{E1} \equiv |p_4/p_2|^2$.
Theory gives $r_{E1}\simeq 1$, while from
the fit it results that $r_{E1}\approx 0.74\pm0.04$.
It is important to stress that the calculation of these RMEs is not
influenced by uncertainties in the two-body currents, since their
values are entirely given by the long-wavelength form of the $E_1$
operator (Siegert's theorem), which has no spin-dependence (for a thorough
discussion of the validity of the long-wavelength approximation
in $E_1$ transitions, particularly suppressed
ones, see Ref.~\cite{Viv00}).  It is therefore
of interest to examine more closely the origin of the above discrepancy.
If the interactions between the $p$ and $d$ clusters are switched off,
the relation $r_{E1} \simeq 1$ then simply follows
from angular momentum algebra.
Deviations of this ratio from one are therefore
to be ascribed to differences induced by the interactions in the
$S$=1/2 doublet and quartet wave functions.
The interactions in these channels do not change
the ratio above significantly.  It should be emphasized that
the studies carried out up until now ignore, in the continuum states, the
effects arising from electromagnetic interactions
beyond the static Coulomb interaction between protons.
It is not clear whether the inclusion of these
long-range interactions, in particular their spin-orbit component,
could explain the splitting between
the $p_2$ and $p_4$ RMEs observed at very low energy.
This discrepancy seems to disappear at 2 MeV~\cite{Viv00}.
\begin{table}[htb]
\caption{Magnitudes of the leading $M_1$ and $E_1$ RMEs
for $pd$ capture at $E_p=40$ keV.}
\label{tab:rmew}

\begin{tabular}{@{}llll}
RME & IA & FULL & FIT \\
\hline
 $|m_2|$ & 0.172  & 0.322 & 0.340$\pm$0.010  \\
 $|m_4|$ & 0.174  & 0.157 & 0.157$\pm$0.007  \\
 $|p_2|$ & 0.346  & 0.371 & 0.363$\pm$0.014  \\
 $|p_4|$ & 0.343  & 0.378 & 0.312$\pm$0.009  \\
\hline
\end{tabular}\\[2pt]
\end{table}

Finally, the doublet $m_2$ RME is underpredicted by theory at the
5 \% level.  On the other hand, the cross section for $n$$d$
capture at thermal neutron energy is calculated
to be 229 $\mu$b with one-body currents and 578 $\mu$b with one- and
many-body currents, using the AV18/UIX Hamiltonian
model~\cite{Viv96}.  This last result is 15 \% larger
than the experimental value (508$\pm$15) $\mu$b~\cite{JBB82}.
Of course, $M_1$ transitions (which induce the $nd$
capture), particularly doublet ones, are significantly
influenced by many-body current contributions.
This is an unsettling state of affairs: on the one hand, theory 
underpredicts the doublet $M_1$ \lq\lq experimental\rq\rq
RME for $pd$ capture (see Table~\ref{tab:rmew}),
while overpredicting, on the other hand, the $nd$
capture cross section, which is dominated by the doublet
$M_1$ transition.

\subsection{The $\alpha d$ Radiative Capture}

Radiative capture of deuterons on $\alpha$ particles is the only process
by which $^6$Li is produced in standard primordial nucleosynthesis
models~\cite{nls}.  There are no direct $\alpha$$d$ capture data
in the energy region relevant for big bang nucleosynthesis (BBN), and
it is therefore crucial to have a reliable theoretical
estimate for the $d(\alpha,\gamma)^6{\rm Li}$ cross section.

In fact, the theoretical description of
the $\alpha d$ capture is particularly
challenging: the S- and P-wave captures are strongly inhibited by
quasi-orthogonality between the initial and final states and by an
isospin selection rule, respectively.  As a result, the dominant
process in all experiments performed to date has been electric
quadrupole ($E_2$) capture from D-wave scattering states.  The small
remaining $E_1$ contribution from P-wave initial states has been
observed at about 2 MeV, but its magnitude has not been successfully
explained by theoretical treatments; it is generally expected to
contribute half of the cross section at 100 keV.  The S-wave capture
induced by $M_1$ has been neglected in most calculations because of the
quasi-orthogonality mentioned above, which makes the associated matrix
element identically zero in two-body treatments of the process.  The
energy dependences of the various capture mechanisms ($E_2$, $E_1$,
$M_1$) are such that even $E_1$ and $M_1$ captures with small amplitudes
may become important at low ($< 200$ keV) energies.  Low-energy
behavior is particularly important for standard BBN: the primordial
$^6$Li yield is only sensitive to the capture cross section between 20
and 200 keV, with the strongest sensitivity at 60 keV~\cite{nb99}. 
\begin{figure}[bth]
\centerline{
\epsfig{file=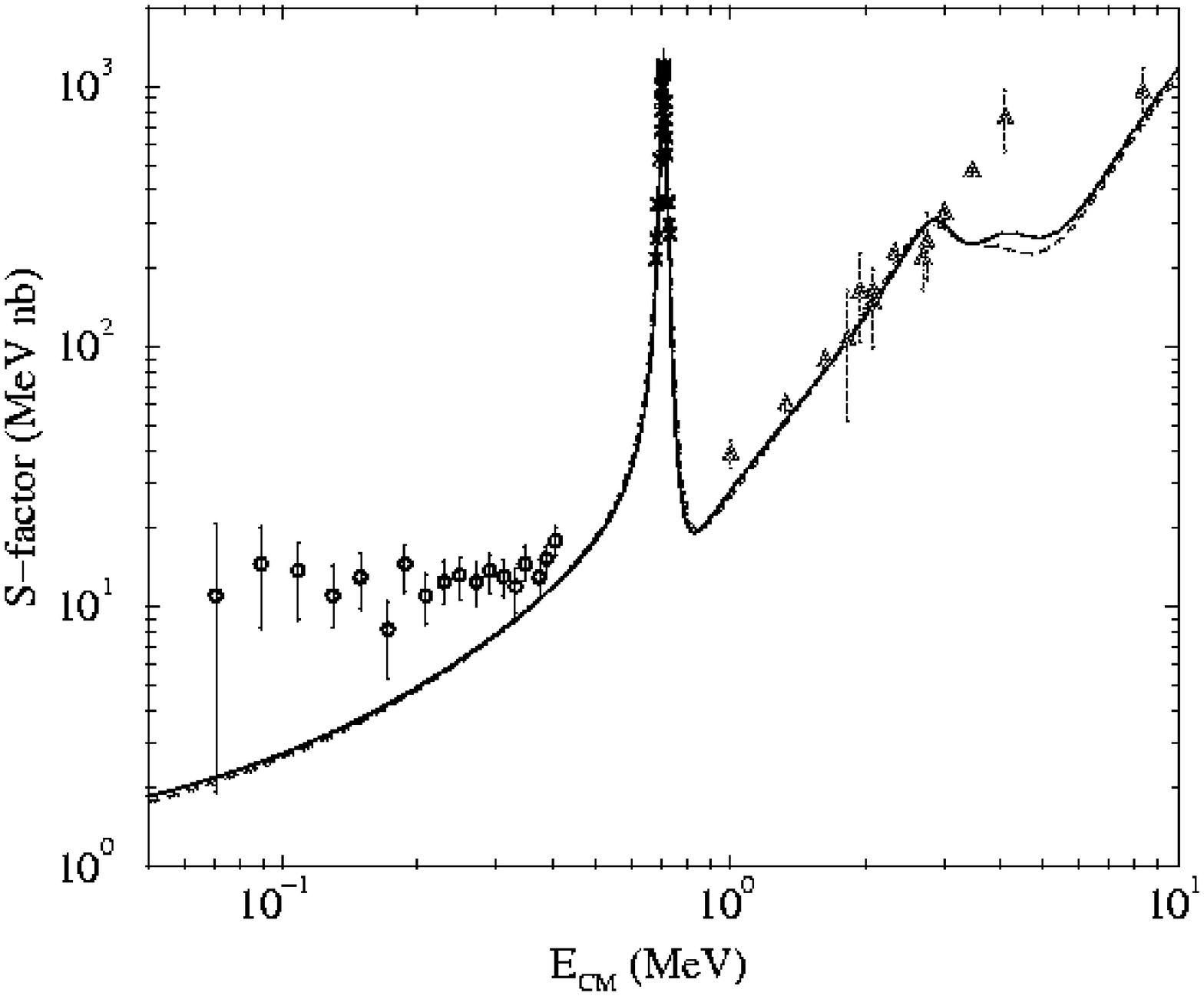,height=3in,width=4in}}
\caption{The calculated $\alpha$$d$ $S$-factor, as obtained
from two distinct $\alpha$$d$ potentials (see text), is compared to available
data.}  
\label{fig:ads}
\end{figure}

Recently, a calculation of the $\alpha d$ capture has been carried out~\cite{NWS00}
using variational Monte Carlo wave functions to describe the initial
$\alpha$$d$ S-, P-, and D-wave scattering states and final $^6$Li
bound-state, and including $M_1$, $E_1$, and $E_2$ transitions.
The resulting $S$-factor is compared with available data in Fig.~\ref{fig:ads}.
While this calculation is not in the same league as the three-body
capture calculations discussed in the previous section or the $p\,^3$He weak
capture calculation presented by Marcucci in these proceedings~\cite{Mar00},
it is, nevertheless, the first attempt to a microscopic description of a six-body
capture process.  At present, the $\alpha$$d$ relative wave function is obtained
from a phenomenological potential, fitted to $\alpha$$d$ elastic scattering data,
and the $^6$Li variational wave function gives an energy that is higher than that
corresponding to separated $\alpha$ and $d$ clusters.  These
aspects of the calculation are clearly unsatisfactory.  For example, the orthogonality
between the $\alpha$$d$ S-wave scattering state and $^6$Li bound state
needs to be enforced artificially.  Some of these limitations can
be easily overcome, for example one could use a Green's function
Monte Carlo wave function for $^6$Li, which does reproduce
the experimental binding energy.  Others, however, are more challenging, one
clearly belonging to this class is the calculation of the $\alpha$$d$ scattering
state from the six-body realistic Hamiltonian.  This is perhaps the central
problem in calculations of capture processes involving systems with $A > 4$. 
\section{Acknowledgments}

I wish to thank J.\ Carlson, J.L.\ Forest, A.\ Kievsky, L.E.\ Marcucci,
K.M.\ Nollett, V.R.\ Pandharipande, S.C.\ Pieper, D.O.\ Riska,
S.\ Rosati, M.\ Viviani, and R.B.\ Wiringa for their many important
contributions to the work reported here.  This work was supported by
DOE contract DE-AC05-84ER40150 under which the Southeastern Universities
Research Association (SURA) operates the Thomas Jefferson National
Accelerator Facility.
\end{document}